\documentclass [12pt]{article}

\begin{document}
\baselineskip=8mm

\begin{center}
\textbf{Experimental studies of light propagation and storage}
\end{center}

\begin{center}
\textbf{in warm atomic gases}
\end{center}

\begin{center}
H. Gao, M. Rosenberry*, J. Wang, H. Batelaan
\end{center}

\begin{center}
\textit{Department of Physics and Astronomy, University of Nebraska -- Lincoln,}
\end{center}

\begin{center}
\textit{116 Brace Lab, Lincoln, NE 68588-0111}
\end{center}

\begin{center}
\textit{*Department of Physics, Sienna College, 515 Loudon Road, Loudonville, NY 12211-1462}
\end{center}

We report on an experimental study of light pulse propagation and
storage in a Rb atomic vapor for different pulse durations,
magnetic fields, and atomic densities, and for two different
isotopes. The results have been analyzed and compared with
previous studies.


\newpage
The resonant interaction of light with three-level $\Lambda $-type
atoms has attracted considerable attention in recent years. Due to
the dramatic change of the index of refraction of the atomic vapor
close to resonance, many interesting effects occur [1-9],
important examples of which are electromagnetically induced
transparency (EIT) [1] and ultraslow group velocity [2,6]. By now
both the EIT effect and the slow group velocity for a pulse of
light have been well studied. With the combination of these two
techniques, light can be stored and stopped with modest absorption
in the atomic ensemble, as has been demonstrated recently [10-19].
The light storage technique may provide a feasible way to realize
a kind of quantum memory for photons [15] and high-efficiency
photon counting [20]. There are a large number of parameters that
affect the practicability of the storage of light technique. The
purpose of this paper is to study the dependence of some of the
essential experimental parameters that are readily accessible. The
specific questions that we address here are the ones we asked
while setting up our experiment: $(a)$ ``How fast should the pulse
be?'', $(b)$ ``How good should the magnetic field shielding be?'',
$(c)$ ``In what temperature range can storage be achieved?'', and
$(d)$ ``Can we use the $^{85}$Rb isotope in stead of $^{87}$Rb?''

$(a)$ Recently theoretical studies show [10,15] that the pulse
spectral width should be contained within the EIT transparency
window to avoid loss and dissipation, which is known as the
adiabatic condition. If the pulse is too short then the atomic
population should not respond to setup EIT within the signal pulse
duration. By reducing the light pulse duration the frequency width
of the pulse can be made increasingly larger than the EIT
transparency window and one can study how much dissipation occurs,
and how the light storage would be affected.

$(b)$ The above information depends on the external magnetic
fields since these conditions can directly change the atomic EIT
properties. On the other hand, nonlinear magneto-optic rotation
(NMOR) can also occur in the resonant interaction of light with
atoms in a $\Lambda $-type atomic structure. NMOR is an optical
rotation effect and can be used in high-precision laser
spectroscopy and magnetometry [5-9]. Typically, optical rotation
induced by an atomic system in steady state is studied. However,
because the control laser is switched, it is important in this
study to distinguish the time-dependent optical rotation effects
from the light storage. This means that observing a time-dependent
signal with the two peaks that look like the typical pre-storage
and post-storage pulse, does not necessarily mean that storage has
been observed. The act of switching on the control laser can
produce a peak that looks like a post-storage signal, even in the
absence of a signal pulse.

$(c) $The EIT also depends on the atomic density. At low Rb cell temperatures the
smaller atomic density makes the EIT signal too weak to be detected. At high
Rb cell temperatures the absorption of the laser pulse is too strong to
allow a signal to be observed.

$(d)$ Finally, by comparing light storage in $^{87}$Rb vapor, to storage in
$^{85}$Rb, one might expect a stronger signal, since the $^{85}$Rb isotope
is more abundant than the $^{87}$Rb isotope.

In this paper, we find that storage of light works for pulse
durations of several $\mu $s to 100 $\mu $s, which in some cases
does satisfy and in other cases does not satisfy the adiabatic
condition. For linear polarization it is not easy to distinguish
the light storage from the optical rotation when an ambient
magnetic field is present. We find that optical rotation occurs at
the time when control light switches on again, which results in a
nontrivial storage signal. To avoid this, good magnetic shielding
(within 2 mG) is necessary. We find the working temperature to
range from 65-90\r{}C. We demonstrate storage in $^{85}$Rb and
determine the optimum detuning for the strongest storage signal.

The experimental setup is schematically shown in Fig.1b, and is
similar to the recent light storage experiment [12, 16-18].
Throughout this study we are using linearly polarization light. An
extended cavity diode laser with $<$ 300 kHz bandwidth is used as
the linearly polarized light source. An acoustic-optical modulator
(AOM) and iris allow for rapid switching of this source. The
linearly polarized laser light can be slightly rotated by a fast
Pockel's cell ($\sim $100ns) to create a weak pulse, which is the
signal field. The polarization of the weak pulse is perpendicular
to that of the remaining light, which serves as the control field.
Our Rb cell is 4 cm long, 2 cm diameter, and contains about 5 Torr
of Helium buffer gas. A solenoid is placed around the Rb cell to
precisely control the static magnetic field along the propagation
direction of the laser beam. The solenoid is enclosed within
double layered magnetic shielding to insure long lifetimes of the
atomic Zeeman coherence. The atomic density can be varied by an
Aerobiax heater cable, which is wrapped around the magnetic
shielding. A polarizing beam splitter (PBS) and two photodiodes
comprise the detection system. The PBS separates the control and
signal fields after the light has passed the Rb cell, and sends
them to the different photodiodes. To create the EIT effect, the
laser frequency is adjusted to the D1 transition for $^{87}$Rb
($\lambda $=794.987 nm), i.e., 5S$_{1 / 2}$, F=2 $ \to $ 5P$_{1 /
2}$, F=1, which is checked by observing the fluorescence and
absorption spectrums. Alternatively the laser is adjusted to the
D1 transition for $^{85}$Rb ($\lambda $=794.984 nm), i.e., 5S$_{1
/ 2}$, F=3 $ \to $ 5P$_{1 / 2}$, F=2. The laser beam diameter is
about $\sim $5 mm with an output power of $\sim $8 mW (2.5 mW
incident on the cell).

We now proceed to describe our experimental results. $(a)$ We
first perform the experiment of slow light propagation and storage
of light for different signal pulse durations. The results are
shown in Fig. 2a and 2b, respectively. We emphasize that two
experimental situations are considered. In the slowing experiment
(where the signal pulse is delayed after propagation through the
Rb cell) the control field is always present, while in the storage
experiment the control field is switched off and on. In the
slowing experiment, the original signal pulse duration ($\Delta
_{pulse})$ and the reference pulse maximum intensity ($I_{ref})$
are measured off-resonance, while the $\Delta '_{pulse}$ and $I'$
are pulse duration and maximum intensity are measured
on-resonance. From Fig. 2a (open circles), we find that the pulse
becomes broader at small pulse durations, which corresponds to a
wider frequency spectral width. In addition, associated with the
pulse broadening is a decrease of the pulse maximum intensity,
which means more absorption occurs (squares). The broadening
effect is nearly eliminated for pulses longer than 30 $\mu $s
duration, while the light intensity still increases slightly with
the pulse duration. The inset of Fig. 2a shows the relation of the
pulse delay time to the pulse durations. It is found that the
delay increases with pulse duration and also saturates after the
characteristic time of 30 $\mu $s. We also found that this
characteristic time increases with higher temperatures.

In the light storage experiment, the time sequence of the control and signal
fields is the same as in Ref. [12] with a storage time of 130 $\mu $s. Light
storage efficiency ($\eta )$ is calculated as the ratio of the maximum light
intensity between post- and pre-storage (or output and input) after
subtracting the leakage from the control beam. The light storage efficiency
increases with pulse durations (Fig. 2b).

$(b)$ When a magnetic field is applied along the laser propagation
direction, the slow light intensity, $I_{s - slow}$, shows a
dispersion curve as a function of the magnetic field (Fig. 3aii,
filled squares), while the storage signal, $I_{s - store}$, shows
three peaks (Fig. 3bi filled squares). The pulse delay at
different magnetic fields is maximum around zero magnetic field
(Fig. 3aiii, left). When we do not send a signal pulse, but only
look at the control light, we see the effect of optical rotation.
The intensity in our signal detector due to optical rotation of
the control field, $I_{c - rotate}$, is given when the control
light is continuously on and reached steady state (Fig. 3ai, open
squares). The optical rotation signal shows a nice symmetric
structure with respect of zero magnetic field for the steady
state. The intensity in our signal detector due to optical
rotation of the control field, $I_{c - rotate}$, is also given at
the time when control light turns on again (Fig. 3bi, open
squares). In this case, the optical rotation signal shows four
peaks. This means that at non-zero magnetic fields signals are
generated that look like storage signals in the absence of a
signal pulse (Fig. 3bii).

$(c)$ The effect of different atomic densities on light slowing and storage are
shown in Fig. 4a and 4b. The total atomic density (with abundances for
$^{85}$Rb of 72{\%} and for $^{87}$Rb of 28{\%}) is estimated from the cell
temperature by using Killian's (semi-empirical) formula [21], $N( /
\mbox{cm}^{3}) = 10^{10.55 - 4132 / T} / (1.38\times 10^{ - 16}T)$, where
$N$ and $T$ are the atomic density (/cm$^{3})$ and cell temperature (K),
respectively. For the slowing experiment an increased pulse delay is
observed for higher atomic densities (Fig. 4a). The inset of Fig. 4a shows
the pulse broadening effect at different temperatures. For the light storage
experiment the efficiency increases for higher atomic densities (Fig. 4b).

We now turn our attention to the discussion of our above results. First we
give the general framework in which our experimental results can be
understood. For our present results it is sufficient to consider the
$\Lambda $-type structure for the linear-linear coupling case. More detailed
theoretical studies on light propagation and storage in three-level systems
can be found in Refs. [15,22,23]. Typically, the susceptibility for a
resonant laser field propagating through an ideal, homogenous EIT medium can
be expressed as [3,15],

\begin{equation}
\label{eq1}
\chi = \kappa \gamma {\frac{{\delta}} {{\Omega _{^{\mbox{c}}}^{2} - \delta
^{2} - i\gamma \delta}} },
\end{equation}

\noindent
where, $\kappa = 3N\lambda ^{3} / 8\pi ^{2}$ is a constant, \textit{$\lambda $} is the resonant
laser wavelength, \textit{$\gamma $} is radiative decay rate of the excited level to the
ground level, and $\delta = \omega - \omega _{0} $ is the laser detuning.
The dispersion relation of light propagation in such a medium,
($k$c/\textit{$\omega $})$^{2}$=1+\textit{$\chi $}, can be found from substituting plane waves in the wave
equation. This can be reduced to

\begin{equation}
\label{eq2}
k\mbox{c} / \omega = 1 + \chi / 2,
\end{equation}

\noindent for small values of the susceptibility near resonance,
where $\omega $ and $k$ are the light circular frequency and wave
number. Generally, the susceptibility is divided into real and
imaginary parts: $\chi = \chi ' + i\chi ''$. For the real part the
index of refraction is $n = 1 + \chi ' / 2$ and for the imaginary
part the absorption coefficient is $\alpha = \omega \chi '' /
2\mbox{c}$. After the light pulse propagates through the atomic
vapor cell of length $L$, the intensity of the pulse is attenuated
by a factor of exp(-2\textit{$\alpha $L}), which as a function of
frequency gives the usual EIT transmission profile. The group
velocity is defined as, $v_{g} = \partial \omega / \partial k$,
which can be obtained by differentiating Eq. (\ref{eq2}) with
respect to $\omega $, $v_{g} = \mbox{c} / (1 + \omega / 2 \cdot
\partial \chi ' / \partial \omega )$. To obtain this relation,
$\chi ' < < \omega \;\partial {\kern 1pt} \chi ' / \partial \omega
$ has been used. This approximation is justified by Eq.
(\ref{eq1}) near resonance. After the light pulse propagates
through the atomic vapor cell of length $L$, its envelope is
delayed compared to free space propagation by a time $T_{g} = L /
v_{g} - L / c = {\frac{{L}}{{c}}} \cdot {\frac{{\omega}} {{2}}}
\cdot {\frac{{\partial \chi '}}{{\partial \omega}} }$.

Optical rotation can be understood by considering the two orthogonal
circularly polarized components of a linearly polarized beam. The two
components couple the two atomic lower states to the upper state in a
$\Lambda $-type atom. Suppose the upper state has m = 0 and two lower states
have m = $\pm $1. In the presence of a magnetic field along the light
propagation direction, the Zeeman effect gives a frequency shift $\pm \mu
_{B} B / \hbar $ to the two lower states, where $\mu _{B} $ is the Bohr
magneton. Thus the two circularly polarized fields experience two different
indices of refraction and optical rotation occurs. The rotation angle,
$\Delta $\textit{$\theta $}, can be defined as [24] $\Delta \theta = 2\pi / \lambda \cdot
[n(\omega _{ +}  ) - n(\omega _{ -}  )] \cdot L / 2$, where $\omega _{\pm}
= \omega _{0} \pm \mu _{B} B / \hbar $ and $\omega _{0} $ is the resonant
circular frequency.

Now we proceed to apply the above general considerations to our results.
$(a)$ Increasing the time duration of the signal pulse leads to a decreasing
spectral pulse width. Starting with a very short pulse would mean that the
spectral pulse width exceeds the EIT window and the size of the output pulse
is very small. The time duration of the output pulse is limited by the
Fourier transform of the EIT window. Conversely, when the time duration of
the signal pulse has become very long, its spectral width is much narrower
than the EIT window. Consequently the pulse is fully transmitted without
broadening. In Lukin and Fleischhauer's paper [3,15], the EIT transparency
window is given by

\begin{equation}
\label{eq3} \Delta \omega _{trans} = (\Omega ^{2} / \gamma ) /
(\kappa kL)^{1 / 2}.
\end{equation}

\noindent In our experiment, the incident laser power on our Rb
cell is 2.5 mW, while 0.25 mW exits after cell with a 5 mm beam
diameter. The Rabi frequency ($\Omega = \sqrt {I\gamma ^{2} /
2I_{s}}  $, $I$ is the laser intensity, $I_{s} = 1.6$mW/cm$^{3}$
is the Rb saturation intensity) associated with these laser beam
powers is estimated to be about 12 MHz and 3.8 MHz, respectively.
The predicted transparency window [Eq. (\ref{eq3})] is thus
bounded by 300 kHz and 30 kHz. The experimental data is fitted to
the Fourier transform of the product of a Gaussian pulse and the
EIT transmission profile. This gives an observed transparency
window width of 50 kHz (Fig. 2a, solid line), i.e., within the
expected range. The same calculation (assuming an EIT transmission
of about 70{\%} [12]) also gives the maximum output intensity
(Fig. 2a, dashed line). Because the pulse is partially outside of
the transparency window its intensity decreases for shorter pulses
(Fig. 2a, squares). We would expect that the intensity would
exponentially increase with the same time constant as the pulse
width exponentially decreased. However, we find that the intensity
still increases after the pulse broadening effect stops. This may
be due to optical rotation at small but non-zero magnetic fields
($<$ 2mG, our detector resolution). This suggestion is supported
by our observation of even larger light intensities when we
increase the magnetic field ($\sim $5 mG, filled squares in Fig.
2a). The optical rotation increases with intensity, which in its
turn increases with pulse duration. We also find that the temporal
delay increases with the pulse durations until the pulse width
matches the EIT window (inset of Fig. 2a). For the light storage
experiment, a similar effect causes the storage efficiency
increase (Fig. 2b) with pulse duration.

$(b)$ Increasing the magnitude of the magnetic field leads to less
EIT [12] and more optical rotation (Fig 3ai). Steady state optical
rotation has been extensively studied before (see, e.g., [7,8]).
Our optical rotation at steady state shows a symmetric structure
with respect to zero magnetic field (Fig. 3ai, open squares). The
detector direction is chosen orthogonal to the electric field
vector of the control light. Optical rotation of this electric
field vector ($\Delta \theta $) into the direction of the detector
explains the symmetry (inset of Fig 3ai). The solid line is
obtained from a density matrix calculation for a three level
atomic $\Lambda $-system. The calculated optical rotation can be
obtained by relating the density matrix coherences to the indices
of refraction as experienced by the circular light components of
the control light. The small asymmetry in the calculation is due
to the use of a slightly different light intensity for $\sigma
^{+}$ and $\sigma ^{-}$.

When an orthogonal linearly polarized signal light is present, the
optical rotation causes the detected signal to be asymmetric with
respect to zero magnetic field as is shown in Fig. 3aii (filled
squares). This contrasts to the situation for circularly polarized
light where a symmetric profile can be observed. The inset of Fig.
3aii illustrates the observed asymmetry. Signal light is generated
by optical rotation from the control light in the Pockel's cell.
Optical rotation due to the Rb vapor ($\Delta \theta $) towards
the detector direction increases the signal, and vice versa. This
feature is also obtained from our model calculation (Fig. 3aii,
solid line).

We find the delay of the signal pulse shows a sharp peak at zero magnetic
field (Fig 3aiii, left). This peak is similar to the usual EIT absorption
profile for circular polarized light (Fig 3aiii, right). This can be
understood by realizing that the EIT absorption rate is proportional to
$\chi ''$, while the time delay is proportional to $\partial \chi ' /
\partial \omega $.

Unlike the steady state, the optical rotation shows some different
features when the control light turns on again. A time spectrum
shows a peak (Fig. 3bii), which we will refer to as
switch-induced-optical rotation (SIR). As a function of magnetic
field this SIR gives four peaks in our experiment (Fig. 3bi, open
squares). This result can again be simulated qualitatively from
our model (inset of Fig. 3bi). Associated with the SIR, the
regenerated storage signal after simply subtracting the SIR shows
three peaks (note that it is still partially from SIR). These
three peaks are due to storage signal and rotation signal. Since
SIR affects the storage signal dramatically, one should be very
careful when interpreting light storage in the presence of a
magnetic field. Our further theoretical analysis shows that SIR is
sensitive to the experimental conditions, such as atomic density,
light intensity, light polarized states, and spin decoherence
time, etc. Even the qualitative features such as the number of
peaks is sensitively dependent on these experimental parameters.
This situation contrasts the case of circular polarization. Then
the optical rotation does not play a role, which gives clean
storage signals.

$(c)$ The time delay of the slowed signal pulse increases with atomic density
(Fig. 4a). This is expected because the expression for $T_{g}$ is
proportional to the atomic density. In addition, for fixed pulse duration,
as the atomic density increases, the EIT transparency window becomes
narrower [3,15]. The slowed pulse thus becomes broader and smaller. This
effect is shown in the inset of Fig. 4a. Larger pulse delays lead in the
storage experiment to larger stored signals (Fig. 4b). For the highest
atomic densities the storage efficiency decreases due to enhanced absorption
from nearby hyperfine levels.

$(d)$ So far, all the above experiments are done in $^{87}$Rb
vapor. Can storage be achieved in the $^{85}$Rb isotope? For this
purpose we adjust our laser frequency to $\lambda $=794.984 nm,
i.e., the 5S$_{1 / 2}$, F=3 $ \to $ 5P$_{1 / 2}$, F=2 transition
for $^{85}$Rb. The hyperfine structure is similar to $^{87}$Rb. By
choosing the operating temperature a little lower than that used
for $^{87}$Rb, we find the light storage signal for $^{85}$Rb,
shown in Fig. 5. The maximum storage efficiency is always lower
than $^{87}$Rb and storage is seen in a somewhat limited
temperature range (60-75\r{} C). We suspect that this is related
to the details of the hyperfine structure. Although the hyperfine
structure is similar, the exact level spacings are different.
Indeed, the storage efficiency as a function of detuning is quite
different for $^{85}$Rb (inset of Fig. 5). Note that our storage
time spectrum shown is taken at a laser frequency -400 MHz detuned
from resonance.

In summary, we have experimentally studied resonant light pulse
propagation and storage in a warm Rb atomic vapor for different
experimental parameters and can answer the four questions raised
in the introduction. We find $(a)$ the light storage signal still
exists at very small pulse duration (several $\mu $s in our
experimental case). However the slowing signal pulse becomes broad
and small. To avoid this effect, the signal spectral bandwidth
should be within the EIT transparency window, which is about 50
kHz in our experiment. The size of the EIT transparency window is
found to depend on the vapor temperature. $(b)$ We have measured
the magnetic field dependence of the slowing and storage signals.
One might expect that the magnetic fields should be limited to the
extent that the Zeeman shifts are within the EIT window. However,
we find it is better to shield the magnetic field to within $\sim$
2 mG (2.8 KHz) for light storage. The reason is that for linearly
polarized light switch-induced-optical rotation (SIR) can mask the
storage signal. $(c)$ Storage could be achieved in a temperature
range of about 65-90\r{}C, which corresponds to the atomic density
of 0.46 x 10$^{12} $cm$^{ - 3}$ to 3 x 10$^{12} $cm$^{ - 3}$. For
the temperature larger than 90\r{} C, the storage signal decreases
due to the absorption. $(d)$ Finally, we demonstrate light storage
for $^{85}$Rb. The light storage efficiency is probably influenced
by the close proximity of other hyperfine lines. This also limits
the storage temperature range of 60-75\r{ }C for $^{85}$Rb.

This work was supported by a Nebraska Research Initiative (NRI) Grant. We
thank B. Williams for his work on the apparatus.

\newpage
\begin{center}
References
\end{center}

[1] See, e.g., S.E. Harris, Phys. Today \textbf{50} (7), 36 (1997).

[2] M.M. Kash, V.A. Sautenkov, A.S. Zibrov, L. Hollberg, G.R. Welch, M.D.
Lukin, Y. Rostovtsev, E.S. Fry, and M.O. Scully, Phys. Rev. Lett.
\textbf{82}, 5229 (1999).

[3] M.D. Lukin, M. Fleischhauer, A.S. Zibrov, H.G. Robinson, V.L.
Velichansky, L. Hollberg, and M.O. Scully, Phys. Rev. Lett. \textbf{79},
2959 (1997).

[4] A.S. Zibrov, M.D. Lukin, L. Hollberg, D.E. Nikonov, M.O. Scully, H.G.
Robinson, and V.L. Velichansky, Phys. Rev. Lett. \textbf{76}, 3935 (1996).

[5] D. Budker, V. Yashchuk, and M. Zolotorev, Phys. Rev. Lett. \textbf{81},
5788 (1998).

[6] D. Budker, D.F. Kimball, S.M. Rochester, and V.V. Yashchuk, Phys. Rev.
Lett. \textbf{83}, 1767 (1999).

[7] I. Novikova, A.B. Matsko, and G.R. Welch, Opt. Lett. \textbf{26}, 1016
(2001).

[8] V.A. Sautenkov, M.D. Lukin, C.J. Bednar, I. Novikova, E. Mikhailov, M.
Fleischhauer, V.L. Velichansky, G.R. Welch, and M.O. Scully, Phys. Rev. A
\textbf{62}, 023810 (2000).

[9] M. Fleischhauer, A.B. Matsko, and M.O. Scully, Phys. Rev. A \textbf{62},
013808 (2000).

[10] M. Fleischhauer and M.D. Lukin, Phys. Rev. Lett. \textbf{84}, 5094
(2000).

[11] O. Kocharovskaya, Y. Rostovtsev, and M.O. Scully, Phys. Rev.
Lett. \textbf{86}, 628 (2001).

[12] D.F. Phillips, A. Fleischhauer, A. Mair, and R.L. Walsworth, and M.D.
Lukin, Phys. Rev. Lett. \textbf{86}, 783 (2001).

[13] C. Liu, Z. Dutton, C.H. Behroozi, and L.H., Hau, Nature (London)
\textbf{409}, 490 (2001).

[14] A.S. Zibrov, A.B. Matsko, O. Kocharovskaya, Y.V. Rostovtsev, G.R.
Welch, and M.O. Scully, Phys. Rev. Lett. \textbf{88}, 103601 (2002).

[15] M. Fleischhauer and M.D. Lukin, Phys. Rev. A \textbf{65}, 022314
(2002).

[16] A. Mair, J. Hager, D.F. Phillips, R.L. Walsworth, and M.D. Lukin, Phys.
Rev. A \textbf{65}, 031802(R) (2002).

[17] M. Kozuma, D. Akamatsu, L. Deng, E.W. Hagley, and M.G. Payne, Phys.
Rev. A \textbf{66}, 031801(R) (2002).

[18] H. Gao, M. Rosenberry, and H. Batelaan, Phys. Rev. A
\textbf{67}, 053807 (2003).

[19] M. Bajcsy, A.S. Zibrov, and M.D. Lukin, Nature (London)
\textbf{426}, 638 (2003).

[20] A. Imamoglu, Phys. Rev. Lett. \textbf{89}, 163602 (2002); D.F.V. James
and P.G. Kwiat, \textit{ibid}. \textbf{89}, 183601 (2002).

[21] T. J. Killian, Phys. Rev. \textbf{27}, 578 (1926).

[22] J.H. Eberly, A. Rahman, and R. Grobe, Laser Phys. \textbf{6}, 69
(1996).

[23] M.G. Payne, L. Deng, C. Schmitt, and S. Anderson, Phys. Rev. A
\textbf{66}, 043802 (2002).

[24] D.A.V. Baak, Am. J. Phys. \textbf{64}, 724 (1996).

\end{document}